\magnification=\magstep1
\openup 2\jot
\def\bo{ { \sqcup\llap{ $\sqcap$} } }
\overfullrule=0pt       

\font\cat=cmr7

\def\zed{Z\hskip -3mm Z }
\def\half{\textstyle{1\over2}}
\def\quarter{\textstyle{1\over4}}
 
\input epsf

     
\newcount\refno
\refno=0
\def\nref#1\par{\advance\refno by1\item{[\the\refno]~}#1}
     
\def\book#1[[#2]]{{\it#1\/} (#2).}

\def\amjph#1 #2 #3.{{\it Am.\ J.\ Phys.\ \bf#1} #2 (#3).}
\def\annph#1 #2 #3.{{\it Ann.\ Phys.\ (N.\thinspace Y.) \bf#1} #2 (#3).}
\def\anndph#1 #2 #3.{{\it Ann.\ der Phys.\ \bf#1} #2 (#3).}
\def\apj#1 #2 #3.{{\it Ap.\ J.\ \bf#1} #2 (#3).}
\def\cmp#1 #2 #3.{{\it Commun.\ Math.\ Phys.\ \bf#1} #2 (#3).}
\def\cqg#1 #2 #3.{{\it Class.\ Quantum Grav.\ \bf#1} #2 (#3).}
\def\foundph#1 #2 #3.{{\it Found.\ Phys.\ \bf#1} #2 (#3).}
\def\grg#1 #2 #3.{{\it Gen.\ Rel.\ Grav.\ \bf#1} #2 (#3).}
\def\ijmpa#1 #2 #3.{{\it Int.\ J.\ Mod.\ Phys.\ \rm A\bf#1} #2 (#3).}
\def\ijmpd#1 #2 #3.{{\it Int.\ J.\ Mod.\ Phys.\ \rm D\bf#1} #2 (#3).}
\def\jmp#1 #2 #3.{{\it J.\ Math.\ Phys.\ \bf#1} #2 (#3).}
\def\jphysa#1 #2 #3.{{\it J.\ Phys.\ \rm A\bf#1} #2 (#3).}
\def\mpla#1 #2 #3.{{\it Mod.\ Phys.\ Lett.\ \rm A\bf#1} #2 (#3).}
\def\mnras#1 #2 #3.{{\it Mon.\ Not.\ R.\ Ast.\ Soc.\ \bf#1} #2 (#3).}
\def\nat#1 #2 #3.{{\it Nature\ \bf#1} #2 (#3).}
\def\ncim#1 #2 #3.{{\it Nuovo Cim.\ \bf#1\/} #2 (#3).}
\def\ncimb#1 #2 #3.{{\it Nuovo Cim.\ \bf#1\/}B #2 (#3).}
\def\np#1 #2 #3.{{\it Nucl.\ Phys.\ \bf#1} #2 (#3).}
\def\npb#1 #2 #3.{{\it Nucl.\ Phys.\ \rm B\bf#1} #2 (#3).}
\def\phrep#1 #2 #3.{{\it Phys.\ Rep.\ \bf#1} #2 (#3).}
\def\pl#1 #2 #3.{{\it Phys.\ Lett.\ \bf#1} #2 (#3).}
\def\pla#1 #2 #3.{{\it Phys.\ Lett.\ \bf#1\/}A #2 (#3).}
\def\plb#1 #2 #3.{{\it Phys.\ Lett.\ \bf#1\/}B #2 (#3).}
\def\pr#1 #2 #3.{{\it Phys.\ Rev.\ \bf#1} #2 (#3).}
\def\prb#1 #2 #3.{{\it Phys.\ Rev.\ \rm B\bf#1} #2 (#3).}
\def\prd#1 #2 #3.{{\it Phys.\ Rev.\ \rm D\bf#1} #2 (#3).}
\def\prl#1 #2 #3.{{\it Phys.\ Rev.\ Lett.\ \bf#1} #2 (#3).}
\def\pcps#1 #2 #3.{{\it Proc.\ Cambs.\ Phil.\ Soc.\ \bf#1} #2 (#3).}
\def\plms#1 #2 #3.{{\it Proc.\ Lond.\ Math.\ Soc.\ \bf#1} #2 (#3).}
\def\prs#1 #2 #3.{{\it Proc.\ Roy.\ Soc.\ \rm A\bf#1} #2 (#3).}
\def\revmod#1 #2 #3.{{\it Rev.\ Mod.\ Phys.\ \bf#1} #2 (#3).}
\def\rprog#1 #2 #3.{{\it Rep.\ Prog.\ Phys.\ \bf#1} #2 (#3).}
\def\sovpu#1 #2 #3.{{\it Sov.\ Phys.\ Usp.\ \bf#1} #2 (#3).}
\def\sovjpn#1 #2 #3.{{\it Sov.\ J.\ Part.\ Nucl.\ \bf#1} #2 (#3).}
\def\sovj#1 #2 #3.{{\it Sov.\ J.\ Nucl.\ Phys.\ \bf#1} #2 (#3).}
\def\jtpl#1 #2 #3.{{\it Sov.\ Phys.\ JETP\ Lett.\ \bf#1} #2 (#3).}
\def\jtp#1 #2 #3.{{\it Sov.\ Phys.\ JETP\ \bf#1} #2 (#3).}
\def\zphys#1 #2 #3.{{\it Zeit.\ Phys.\ \bf#1} #2 (#3).}

\hbox{ }
\rightline {DTP/97/1}
\rightline {gr-qc/9701014}
\vskip 1truecm
 
\centerline{\bf COSMIC STRINGS IN DILATON GRAVITY}
\vskip 1truecm
 
\centerline{Ruth Gregory \& Caroline Santos\footnote{$^*$}
{On leave from: 
Departmento de F\'\i sica da Faculdade de Ci\^encias da Universidade do Porto,
Rua do Campo Alegre 687, 4150-Porto, Portugal}}
\vskip 2mm
\centerline{ \it Centre for Particle Theory, }
\centerline{\it University of Durham, Durham, DH1 3LE, U.K.}
 
\vskip 4mm
\centerline{\cat ABSTRACT}
\vskip 4mm
 
{
\leftskip 10truemm \rightskip 10truemm

\openup -1 \jot

We examine the metric of an isolated self-gravitating abelian-Higgs
vortex in dilatonic gravity for arbitrary coupling of the vortex
fields to the dilaton. We look for solutions in both massless and
massive dilaton gravity. We compare our results to existing metrics
for strings in Einstein and Jordan-Brans-Dicke theory. We explore the
generalization of Bogomolnyi arguments for our vortices and
comment on the effects on test particles.
 
\openup 1\jot
}
 
 
 
\vfill\eject
\footline={\hss\tenrm\folio\hss}
 
\noindent{\bf 1. Introduction.}

Topological defects and other soliton structures have a wide application
to many areas of physics. 
Cosmologists are interested in  
defects as possible sources for the density perturbations which seeded
galaxy formation. String theorists are interested in defects
not only as solutions of the low energy effective action, but
as true solitons in the full non-perturbative theory which are required
for consistency and inter-relation of the full spectrum of string
theories.

A topological defect is a discontinuity in the vacuum,
and in conventional field theory can be classified according 
to the topology of the vacuum manifold of
the particular field theory being used to model the physical set up: 
disconnected vacuum manifolds give domain walls, non-simply connected
manifolds, strings, and manifolds with non-trivial 2- and 3-spheres
give monopoles and textures respectively. In this paper, we are concerned with
defects associated with non-simply connected vacuum manifolds: cosmic
strings[1]. The gravity of cosmic strings within the context of Einstein
theory has been well explored, both in the case of `model' strings, where the
core of the string is modelled by a simplified energy-momentum tensor[2],
and in the case of the fully coupled Einstein-Abelian-Higgs system[3,4]; with
the result that the spacetime of a self-gravitating local string is found to be
generically conical, with the angular deficit given by $8\pi G\mu$, $\mu$
being the energy per unit length of the vortex.

It seems likely however, that gravity is not given by the Einstein action,
at least at sufficiently high energy scales, and the most promising alternative
seems to be that offered by string theory, where the gravity becomes
scalar-tensor in nature[5]. Scalar-tensor gravity is not new, it was
pioneered by Jordan, Brans and Dicke[6], who sought to incorporate Mach's
principle into gravity. The implications of such actions on general
Friedmann-Robertson-Walker cosmological models have been well explored[7,8],
however, the implications for theories of structure formation have not been so
well studied. Broadly speaking, there are two views on explaining structure
formation -- inflation or defects, the latter consisting of two subsets: cosmic
string or texture[9] induced perturbations. 
While there is little to choose between
these from the particle physics or large scale structure point of view, the
implications of each of these theories for the perturbations of the microwave
background are distinct. However, calculations on the microwave background
multipole moments do assume Einstein gravity[10], therefore it is interesting to
question whether these conclusions are still valid in the context of
scalar-tensor gravity. Even if the dilaton acquires a mass
at a fairly high energy scale (with respect to the recombination
temperature of the universe that is), at the core of
a defect symmetry is restored and the physics
is determined by the GUT scale, at which the dilaton might
have rather different properties, impacting back on
the cosmic microwave background.

Calculations involving radiation from a cosmic string network generally make
use of a ``worldsheet-approximation'' in which the string is treated as an
infinitesimally thin source which moves according to, and has an energy
momentum tensor appropriate for a two-dimensional worldsheet governed by the
Nambu action. That this action is appropriate for the local string has been
convincingly argued in the absence of gravity[11,12], but as yet no proof exists 
in the presence of gravity. This is generally believed to be  related to the
problems of using distributional sources of codimension greater than one in
general relativity[13]. Nonetheless, the fact that the self-gravitating
infinite local vortex has a relatively small effect on spacetime lends credence
to the worldsheet approximation for the string.

In the presence of a dilaton, the worldsheet approximation may no
longer be appropriate. If the dilaton is massless, there is no reason to
expect that the string will not have a long range effect on the dilaton, and
even if the dilaton is massive, it introduces an additional length scale 
which may still have significant impact.

In this paper, we take a modest step towards resolving this issue by
examining the gravi-dilaton field of a self-gravitating cosmic string in
dilaton gravity. We consider a reasonably general form for the interaction
with the dilaton, assuming that the abelian-Higgs lagrangian couples to the
dilaton via an arbitrary coupling, $e^{2a\phi}{\cal L}$, in the string frame.
We consider both massive and massless dilatons. Our results for the massless
dilaton are very similar to those of Gundlach, Ortiz and others [14], 
who considered
cosmic strings in JBD theory. For the massive dilaton we find that, apart from
an intermediate annular region, the long-range structure of the string is as
for Einstein gravity, as might be expected. 
The main exception to this qualitative and expected picture is that for
a special value ($a = -1$) of the coupling of the dilaton to the fields which
constitute the vortex the dilaton effectively decouples from
the string, showing little or no reaction to its presence. This occurs
independent of whether the dilaton is massive, and independent of the
specifics of the U(1) model, i.e.\ whether it is type I, II, or 
supersymmetric.

The layout of the paper is as
follows: In the next section we review the Nielsen-Olesen vortex in the
abelian-Higgs model. In section three we derive the main results of this
paper, namely the gravitational and dilaton fields for the self-gravitating
vortex in both massless and massive dilatonic gravity. In section four we
consider Bogomolnyi bounds for the string, and show that these can only be
saturated in the special case $a = -1$. In this case, the dilaton effectively
decouples from the string. In section five we consider the motion of test
particles in the background of the string, and in section six we conclude. 

\vskip 5mm
 
\noindent{\bf 2. The Abelian Higgs Vortex.}
 
\vskip 2mm

We start by briefly reviewing the U(1) vortex in order to establish notation
and conventions. We take the abelian Higgs lagrangian
$$
{\cal L}[\Phi ,A_a] = D_{a}\Phi ^{\dagger}D^{a}\Phi -
{1\over 4}{\tilde F}_{ab}{\tilde F}^{ab} - {{\lambda  }\over 4 }
(\Phi ^{\dagger} \Phi - \eta ^2)^2
\eqno (2.1)
$$
where $\Phi$ is a complex scalar field, $D_a = \nabla _{a} + ieA_{a}$
is the usual gauge covariant derivative, and
${\tilde F}_{ab}$ the field strength associated with $A_a$. 
We use units in which $\hbar=c=1$ and a mostly minus signature.
For cosmic
strings associated with galaxy formation $\eta \sim 10^{15}$GeV.
 
We rewrite the fields in a way which makes manifest 
the physical degrees of freedom of the model:
$$
\eqalignno{
\Phi (x^{\alpha}) &= \eta  X (x^{\alpha}) e^{i\chi(x^{\alpha}) } &(2.2a) \cr
A_{a} (x^{\alpha}) &= {1\over e}
\bigl [ P_{a} (x^{\alpha}) - \nabla _{a} \chi (x^{\alpha}) \bigr ]\; .
&(2.2b) \cr
}
$$
where $X, \; \chi $ and $P_{a}$ are now real.
In terms of these new variables, the lagrangian
and equations of motion become
$$ 
{\cal L} = \eta ^2 \nabla _{a}X \nabla ^{a}X
+ \eta^2X^2P_{a}P^{a} - {1\over{4e^2}} F_{ab} F^{ab}
-{ {\lambda \eta ^4}\over 4} (X^2 -1)^2 
\eqno(2.3)
$$
$$
\eqalignno{
\bo X
-  P_{a}P^{a}X + {\lambda   \eta  ^2\over 2} X(X^2 -1) &= 0
&(2.4a) \cr
\nabla _{a}F^{a b} + 2 e^2 \eta^2 X^2 P^{b} &=0\; .
&(2.4b) \cr }
$$
Thus $P_{b}$ is the massive vector field in the broken symmetry
phase, $F_{ab} = \nabla_a P_b - \nabla_b
P_a$ its field strength, and $X$ the
residual real scalar field with which it interacts. $\chi$
is not in itself a physical quantity, however, it can
contain physical information if it is non-single valued, in
other words, if $ 
\oint \nabla_a \chi dx^a  = 
{{2\pi}} n $ for  some $ n \in $ {\rm \zed}.
Continuity then demands (in the absence of non-trivial spatial
topology) that $X=0$ at some point on any surface spanning the loop 
- this is the
locus of the vortex. Thus the true physical content of this model is contained
in the fields $P_a$ and $X$  plus boundary conditions on $P_a$  and $X$
representing vortices.
 
The simplest vortex solution is the Nielsen-Olesen (NO) vortex[11], an infinite,
straight static $n=1$ solution with cylindrical symmetry.
In this case, we can choose a gauge in which
$$
\Phi = \eta X_0(R)e^{i\phi} \;\;\; ; \;\;\;
A_a = {1\over e} \left [P_0(R) - 1 \right ]\nabla_a\phi 
\eqno (2.5)
$$
where $R=\sqrt{\lambda}\eta r$, in cylindrical polar coordinates.
The equations for $X_0$ and $P_0$ from (2.4) are
$$
\eqalignno{
-X_0  '' - {{X_0  '}\over R} + {{P_0 ^2 X_0 }\over {R^2}}
+ {\textstyle{1\over 2}} X_0  (X_0 ^2 -1) &= 0 &(2.6a) \cr
-P_0  '' + {{P_0  '}\over R} + {{\beta^{-1}}} X_0 ^2 P_0 &=
0 &(2.6b) \cr }
$$
where a prime denotes ${d\over dR}$, and
${\beta} = \lambda/2e^2 = m^2_{X}/m^2_{P}$ is the Bogomolnyi
parameter[15] (${\beta}=1$ corresponds to the vortex
being supersymmetrizable). Note that in these rescaled
coordinates, the string has width of order unity.  This
string has winding number one; for winding number $N$, we
replace $\chi$ by $N\chi$, and hence $P$ by $NP$.
Figure 1 shows the Nielsen-Olesen solutions for $X$ and $P$ for
a $\beta=1$ winding number one string.
\midinsert \vskip -5truecm \hskip 1truecm \epsfysize=18truecm
\epsfbox{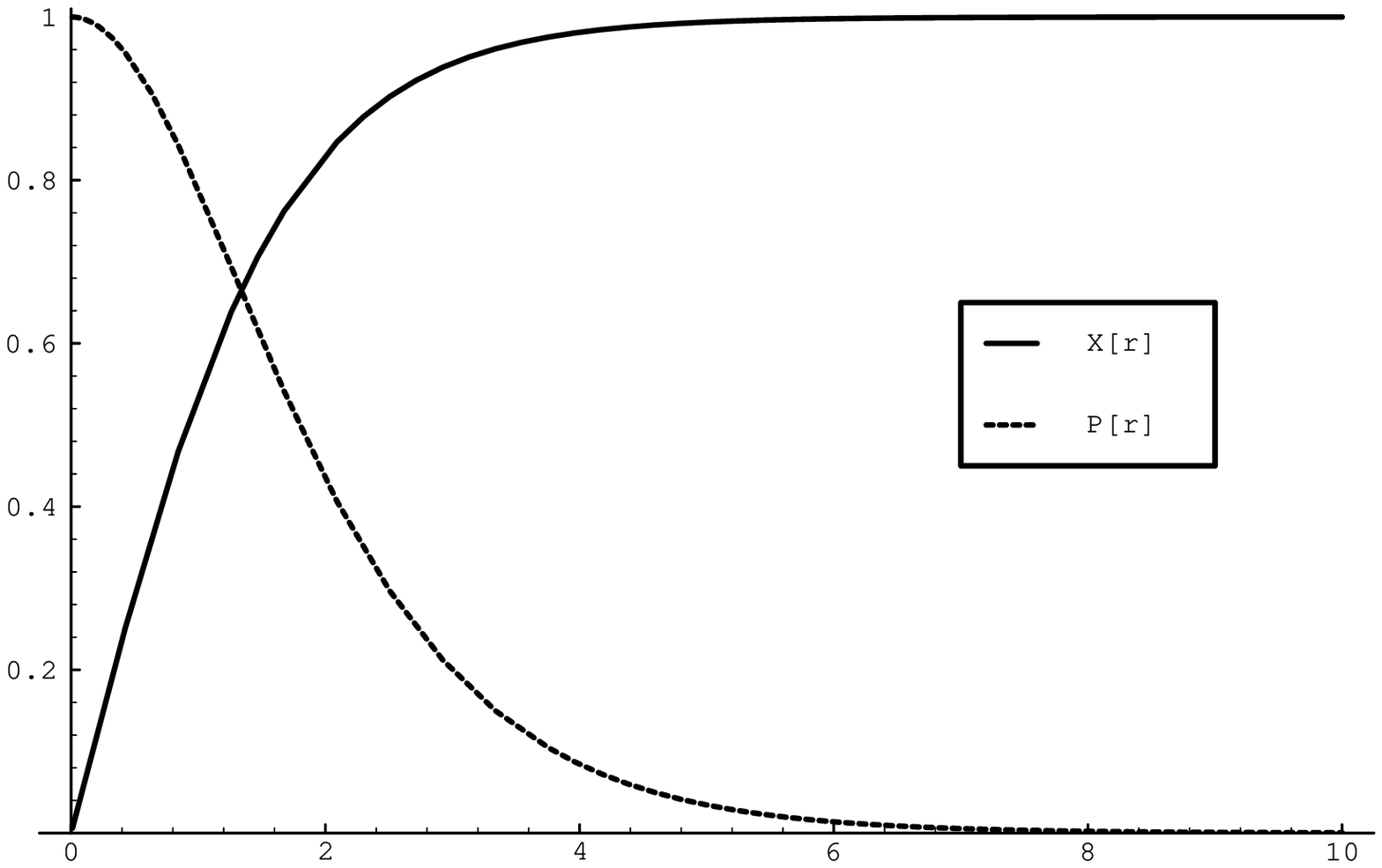} \vskip -5 truecm \hskip 2 truecm
\vbox{ \hsize=11.5 truecm
\noindent {{\cat FIGURE (1):} $X$ and $P$ for a $\beta=N=1$ vortex.  }}
\endinsert

It is also useful to briefly review the self-gravitating NO vortex in Einstein
gravity, as much of the formalism can be used directly in the next section.
To include the self-gravity of the string, we require a metric which 
exhibits the symmetries of the source, namely, translational invariance
along its length and rotational invariance around the core, i.e.~cylindrical
symmetry. The general cylindrically symmetric metric was given
by Thorne [16]
$$
ds ^2 = e ^{2( \gamma - \psi )}(dt ^2 - dr ^2 ) -
e ^{2 \psi }dz ^2 -  {\tilde \alpha} ^2 e ^{-2 \psi } d \phi ^2 
\eqno (2.7)
$$
(where $\gamma, \ \psi, {\tilde \alpha}$ are independent of $z, \phi$).
The string couples to this metric via its energy-momentum tensor
$$
G_{ab} = 8\pi G T_{ab} = 8\pi G\left [ 2\eta^2 \nabla_a X\nabla_b X +
2\eta^2X^2P_a P_b -{2\beta \over \lambda } F_{ac}F_b^{\ c} -
{\cal L}g_{ab}\right ] 
\eqno (2.8) 
$$
which can be seen to be boost invariant ($T^0_0 = T^z_z$). This in turn
implies that for static metrics, $\gamma = 2\psi$, which we will assume
from now on. Note that the expressions (2.7) and (2.8) appear
in unrescaled coordinates, it proves to be convenient to rescale
the coordinates so that the string width is of order unity, as in (2.6). 
To do this  we set
$R = \sqrt{\lambda}\eta r$ as before, $\alpha = \sqrt{\lambda}\eta
{\tilde \alpha}$, and write the rescaled version of the
energy and stresses ${\hat T}_{ab} = T_{ab}/(\lambda\eta^4)$ 
$$
\eqalignno{
{\hat T}^0_0 &= {\cal E} = e ^{-\gamma} X ^{\prime 2} + {e^{\gamma} X^2
P^2 \over {\alpha} ^2} + {\beta P ^{\prime 2}  \over {\alpha} ^2 }+
(X^2 -1) ^2 / 4  & (2.9a) \cr 
{\hat T}^R_R &= - {\cal P}_R = - e ^{-\gamma } X ^{\prime 2} +
{e^{\gamma} X^2 P^2 \over {\alpha} ^2} - {\beta P ^{\prime 2} \over
{\alpha} ^2} + (X^2 -1) ^2 / 4 & (2.9b) \cr 
{\hat T}^\theta_\theta &=  -{\cal P}_\theta = e ^{-\gamma } X ^{\prime 2} -
{e^{\gamma} X^2 P^2 \over  {\alpha} ^2} - {\beta P ^{\prime 2} 
\over {\alpha} ^2} + (X^2 -1) ^2 / 4 \;\;\; & (2.9c) \cr 
{\hat T}^z_z &= - {\cal P}_z = {\hat T}^0_0 .& (2.9d)\cr}
$$
The Einstein equations can then be read off as [16]
$$
\eqalignno{
\alpha'' &= -\epsilon\alpha e^\gamma ({\cal E} - {\cal P}_R) & (2.10a) \cr
(\alpha\gamma')' &= \epsilon \alpha e^\gamma ({\cal P}_R + {\cal P}_\theta)
&(2.10b) \cr 
\alpha'\gamma' &= {\alpha \gamma^{\prime 2} \over 4} + \epsilon\alpha
e^\gamma {\cal P}_R & (2.10c) \cr
}
$$
where $\epsilon=8\pi G\eta^2$ is the gravitational strength of the string.
Also for future reference, the Bianchi identity gives
$$
{\cal P}' _{R} + ({\cal P} _{R} -  {\cal P}_\theta) ({\alpha '\over \alpha}
- {\gamma'\over2}) + \gamma '{\cal P} _R + \gamma '{\cal E}= 0. \eqno (2.11)
$$
 
To zeroth order (flat space)
$$
\alpha = R \;\; , \;\;\; \psi=\gamma=0 \;\; , \;\;\; X=X_0\;\;,\;\;\; P=P_0,
\eqno (2.12)
$$
and (2.11) gives
$$
(R{\cal P}_{_0R})' = {\cal P}_{_0\theta}
\eqno (2.13)
$$
 
To first order in $\epsilon=8\pi G\eta^2$ the string metric is given by[3,4]
$$
{\alpha} = \left [1- \epsilon \int _0 ^R
R ({\cal E}_{_0} - {\cal P}_{_0R})d R \right ] R +
\epsilon \int _0 ^R  R ^2 ({\cal E}_{_0} - {\cal P}_{_0 R})dR,
\eqno (2.14a)
$$
$$
\gamma = \epsilon \int _0 ^R R {\cal P}_{_0 R} dR .
\eqno (2.14b)
$$
where the subscript zero indicates evaluation in the flat space limit.
Note that when the radial stresses
do not vanish, there is a scaling between the
time, $z$ and radial coordinates for an observer at infinity and those for an
observer sitting at the core of the string[4]. The only case in which these
stresses do vanish is when ${\beta}=1$. In this case
the field equations reduce to 
$$
\eqalign{
X' &= XP/\alpha \cr
P' &= {\half} \alpha (X^2 -1) \cr
\alpha ' &= 1 - \epsilon [ (X^2-1)P +1] \cr
\gamma&=0 \cr
}
\eqno(2.15)
$$
a first order set of coupled differential equations as one might expect from
the fact that the solution is supersymmetrizable.
 
We conclude this section by demonstrating the asymptotically conical nature of
the corrected metric. Note that since the string functions $X$ and $P$
rapidly fall off to their vacuum values outside the core, the integrals 
in (2.14) rapidly converge to their asymptotic, constant, values. Let
$$
\epsilon \int _0 ^R
R ({\cal E}_{_0} - {\cal P}_{_0R})d R = A, \;\;\;\;
\epsilon \int _0 ^R R^2 ({\cal E}_{_0} - {\cal P}_{_0R})d R = B
\;\;\;\;\;{\rm and} \;\;\;\;\;
\epsilon \int _0 ^R R {\cal P}_{_0 R} = C
\eqno (2.16)
$$
then the asymptotic form of the metric is
$$
\eqalign{
ds^2 &= e^C[dt^2-dr^2-dz^2] - r^2 (1-A+B/\sqrt{\lambda}\eta r)^2 e^{-C}
d\theta^2\cr 
&= d{\hat t}^2 - d{\hat r}^2 - d{\hat z}^2 - {\hat r}^2 (1-A)^2
e^{-2C} d\theta^2 \cr}
\eqno (2.17)
$$
where 
$$
{\hat t} = e^{C/2}t,\ {\hat z} = e^{C/2}z,\  {\hat r} =
e^{C/2}(r + B/(1-A))
\eqno (2.18)
$$  
This is seen to be conical with a deficit
angle 
$$
\Delta = 2\pi (A+C) = 2\pi \epsilon\int R {\cal E}_{_0} dR
= 16 \pi^2 G \int r T^0_0 dr = 8\pi G\mu
\eqno (2.19)
$$
where $\mu$ is the energy per unit length of the string.
Notice that the deficit angle is independent of the radial stresses,
but that there is a red/blue-shift of time between infinity and the
core of the string if they do not vanish. Now let us examine the behaviour of
the string with a dilaton present.
 
\vskip 5mm

\noindent{\bf 3. Cosmic strings in dilaton gravity.}

\vskip 2mm

We are interested in the behaviour of the isolated string metric (2.7)
when the gravitational interactions take a form typical of low
energy string theory [5]. In its most minimal form, string gravity replaces
the gravitational constant, $G$ by a scalar field, the dilaton in a  rather
analogous fashion to that of Jordan, Brans and Dicke who were motivated by
Mach's principle. We take an empirical approach to cosmic strings in this
background theory, not concerning ourselves with the origin of the fields that
form the vortex, but inputting `by hand' the abelian-Higgs lagrangian (2.1).
To take account of the (unknown) coupling of the cosmic string to the 
dilaton, we choose
$$
{\hat S} = \int  d^4 x \sqrt{-{\hat g}} \left [ e^{-2\phi} \left ( -{\hat R} -
4({\hat \nabla}\phi)^2 - {\hat V}(\phi) \right ) + e^{2a\phi} {\cal L} \right ]
\eqno (3.1)
$$
where ${\cal L}$ is as in (2.3). This action is written in terms of the
string metric, i.e.~the metric which appears in the string sigma model. 
It proves useful to instead write the action in terms of the 
``Einstein'' metric, which is defined via
$$
g_{ab} = e^{-2\phi} {\hat g}_{ab}
\eqno (3.2)
$$
in which the gravitational part of the action appears in the 
more familiar Einstein form:
$$
S = \int  d^4 x \sqrt{-g} \left [ 
- R + 2 (\nabla\phi)^2 - V(\phi)
+ e^{2(a+2)\phi} {\cal L} \{X,P,e^{2\phi}g \} \right ]
\eqno(3.3)
$$
where $V(\phi ) = e^{2\phi}{\hat V}(\phi)$. Note however that this complicates
the matter part of the lagrangian -- a factor of $e^{-2\phi}$ being picked up
each time ${\hat g}^{ab}$ is used:
$$
T_{ab} = 2{\delta {\cal L}[X,P,e^{2\phi}g]\over \delta g^{ab}} = 2\eta^2
e^{-2\phi} [ \nabla_aX\nabla_bX + X^2 P_aP_b] - {2\beta\over\lambda} e^{-4\phi}
F_{ac}F_b^{\ c} - {\cal L} g_{ab}
\eqno(3.4)
$$
The ``Einstein'' equations are now
$$
G_{ab} = {\half} e^{2(a+2)\phi} T_{ab} + S_{ab}
\eqno (3.5)
$$
where
$$
S_{ab} = 2 \nabla_a\phi\nabla_b\phi +{\half} V(\phi)g_{ab} - (\nabla \phi)^2
g_{ab}
\eqno (3.6)
$$
represents the energy-momentum of the dilaton, which has as its equation of
motion
$$
\bo \phi = {\quarter} {\partial V\over\partial\phi} + {(a+1)\over2}
e^{2(a+2)\phi} {\cal L}[X,P,e^{2\phi}g] + e^{2(a+2)\phi} \left [
{\beta\over2\lambda} F^2 e^{-4\phi} - {\lambda\eta^2\over8}(X^2-1)^2 \right ]
\eqno (3.7)
$$

As before, we choose the Thorne metric (2.7), with $\gamma=2\psi$:
$$
ds^2 = e^\gamma(dt^2-dz^2-dr^2) - {\tilde\alpha}^2e^{-\gamma} d\theta^2
\eqno (3.8)
$$
We will also rescale the coordinates again so that the string width is of order
unity, $R=\sqrt{\lambda}\eta r$, $\alpha = \sqrt{\lambda}\eta {\tilde\alpha}$,
and the rescaled modified energy-momentum we redefine as:
$$
\eqalignno{
{\cal E} &= e^{2(a+2)\phi} \left [ e^{-2\phi} \left (
e ^{-\gamma} X ^{\prime 2} +{e^{\gamma} X^2 P^2 \over {\alpha} ^2}\right ) +
e^{-4\phi}{\beta P ^{\prime 2}  \over {\alpha} ^2 }+ (X^2 -1) ^2 / 4 
\right ]  & (3.9a)\cr  
{\cal P}_R &= e^{2(a+2)\phi} \left [ e^{-2\phi} \left (
e ^{-\gamma } X ^{\prime2} - {e^{\gamma} X^2 P^2 \over {\alpha} ^2}\right ) +
e^{-4\phi}{\beta P ^{\prime 2} \over {\alpha} ^2} - (X^2 -1) ^2 / 4 
\right ] & (3.9b) \cr 
{\cal P}_\theta &= e^{2(a+2)\phi} \left [ e^{-2\phi} \left (
-e ^{-\gamma } X ^{\prime 2} +{e^{\gamma} X^2 P^2 \over\alpha^2} \right ) + 
e^{-4\phi} {\beta P ^{\prime 2}  \over {\alpha} ^2} - (X^2 -1) ^2 / 4 
\right ] & (3.9c) \cr }
$$
In terms of these variables, the full equations of motion for the gravitating
vortex in dilaton gravity are
$$
\eqalignno{
\alpha'' &= -\alpha e^\gamma {\tilde V}(\phi) 
-\epsilon\alpha e^\gamma ({\cal E} - {\cal P}_R) & (3.10a) \cr 
(\alpha\gamma')' &= -\alpha e^\gamma {\tilde V}(\phi)  + \epsilon\alpha
e^\gamma ({\cal P}_R + {\cal P}_\theta) &(3.10b) \cr 
\alpha'\gamma' &= -{\half} \alpha e^\gamma {\tilde V}(\phi) + {\alpha
\gamma^{\prime 2} \over 4} + \alpha \phi^{\prime2} +
\epsilon\alpha e^\gamma {\cal P}_R & (3.10c) \cr 
(\alpha \phi')' &= {\alpha e^\gamma\over4} {\partial{\tilde
V}\over\partial\phi} + \epsilon {(a+1)\over2} \alpha e^\gamma{\cal E} -
{\half}\epsilon \alpha e^\gamma ({\cal P}_R + {\cal P}_\theta)
& (3.10d) \cr
{1\over \alpha} (\alpha X')' &= -2(a+1) X'\phi' + {XP^2\over \alpha^2
}e^{2\gamma} + {\half} X(X^2-1) e^{\gamma+2\phi} & (3.10e) \cr
\alpha\left ({P'\over\alpha} \right )' &= -\gamma'P' - 2a\phi'P' + \beta^{-1}
X^2Pe^{\gamma+2\phi} & (3.10f) \cr
}
$$
where $\epsilon = \eta^2/2$ now defines the gravitational strength of the
string, and ${\tilde V} = V/\lambda \eta^2$ represents the dilaton potential in
units natural to the vortex.
The Bianchi identity (2.11) becomes
$$
\epsilon (\alpha e^\gamma {\cal P}_R)' = 
\epsilon \alpha'e^\gamma {\cal P}_\theta + {\half}
\epsilon \alpha\gamma'e^\gamma [ {\cal P}_R - {\cal P}_\theta - 2{\cal E} ]
-\alpha'\phi^{\prime2} - (\alpha \phi^{\prime2})'
+ {\half} \alpha e^\gamma \phi' {\partial V \over \partial \phi}
\eqno (3.11)
$$

We start by examining the case $V(\phi)\equiv0$, i.e.\ a massless dilaton, as
this ought to be qualitatively the same as a cosmic string in Brans-Dicke
gravity.

\vskip 3mm

\noindent{\it 3.1  Massless dilatonic gravity.}

\vskip 2mm

In the case that the dilaton is massless the equations (3.10) are rather
reminiscent of the  pure Einstein gravity vortex (2.10), however, there is one
crucial difference - the constraint equation (3.10c) now contains an $\alpha
\phi^{\prime2}$ term, and unless $a = -1$, $\alpha \phi'$ will definitely be
nonzero. In order to explore this solution, let us first consider the ``wire
approximation'', namely
$$
\alpha e^\gamma {\cal E}(R) = {\hat\mu} \delta(R) \;\;\; ;\;\;\;\;
{\cal P}_R = {\cal P}_\theta =0
\eqno (3.12)
$$
where ${\hat\mu}= \mu/4\pi\epsilon$ represents the energy 
per unit length of the cosmic string in units
natural to the vortex, and is of order unity. (Recall that $\epsilon$ sets the
gravitational strength of the string.) In this case, eqns.(3.10) are readily
integrated to give
$$
\eqalignno{
\alpha(R) &= (1 - \epsilon{\hat\mu})R & (3.13a) \cr
\gamma(R) &= 0 & (3.13b) \cr
\phi(R) &= {\epsilon{\hat\mu} (a+1)\over 2(1 - \epsilon{\hat\mu})}
 \ln R & (3.13c) \cr
}
$$
but now we find a contradiction -- the constraint (3.10c) is no longer
satisfied unless $a = -1$. It is worth examining what has gone wrong here. The
wire model is an approximate version of the stress-energy tensor which usually
works well in Einstein gravity since the integral
$
\int_0^\infty \alpha e^\gamma ({\cal P}_R + {\cal P}_\theta) = 0
$,
which is no longer necessarily true in the presence of the dilaton. A
Bogomolnyi solution in flat space or Einstein gravity has the property that
${\cal P}_R = {\cal P}_\theta\equiv0$, therefore the fact that we cannot
consistently use the wire approximation for these variables (unless $a = -1$)
is an indication that a Bogomolnyi argument cannot exist unless $a = -1$.

Instead, let us examine consistent vacuum solutions to (3.10) which should
represent asymptotic spacetimes for the string. Setting ${\cal E} = 
{\cal P}_R = {\cal P}_\theta = 0$ in (3.10) gives as the general solution:
$$
\eqalignno{
\alpha &= dR+b & (3.14a) \cr
\gamma &= \gamma_0 + {c\over d}\ln (dR+b) & (3.14b) \cr
\phi &= \phi_0 + {f\over 2d}  \ln (dR+b) & (3.14c) \cr
}
$$
where $f = \pm\sqrt{4dc-c^2}$ from (3.10c). This gives a Levi-Civita[17]
solution for the metric. (Note that if $\phi$ is constant, we have $c=0 $ or
$4d$, corresponding to the vacuum general relativistic solutions.) 
The constants $b,c,d,f$ are given by integrating (3.10) and to order
$\epsilon$ are 
$$
d = 1-A, \;\; b = B, \;\; c = 0, \;\; f = {\half}(a+1) (A+C) =
{\half}(a+1)\epsilon{\hat\mu}
\eqno (3.15)
$$
where $A,B,C$ are given in (2.16). We can therefore see that $c$ cannot remain
zero, and to order $\epsilon^2$, $c = {\quarter} (a+1)^2 
\epsilon^2{\hat\mu}^2$. So,
unlike the Einstein self-gravitating vortex, the dilaton vortex for $a \neq -1$
has a strong gravitational effect far from the core, albeit an $O(\epsilon^2)$
one:
$$
\eqalign{
ds^2 &\approx {\hat r}^{(a+1)^2\epsilon^2{\hat\mu}^2/4}
(d{\hat t}^2 - d{\hat r}^2
-d{\hat z}^2) 
- (1-\epsilon{\hat\mu})^2{\hat r}^{2 -(a+1)^2\epsilon^2{\hat\mu}^2/4}
d\theta^2 \cr
e^{2\phi} &\approx e^{2\phi_0} {\hat r}^{(a+1)\epsilon{\hat\mu}/2}\cr
}
\eqno(3.16)
$$
where ${\hat r}$ etc.\ were defined in (2.18) and we have set
$(\sqrt{\lambda}\eta)^{(a+1)^2\epsilon^2{\hat\mu}^2/4} (\simeq
\epsilon^{-\epsilon^2}) \simeq 1$. This metric agrees with Gundlach and
Ortiz[14], who derived the metric for a Jordan-Brans-Dicke cosmic string. In
the string frame,
$$
d{\hat s}^2= e^{2\phi}ds^2 = {\hat r}^{{(a+1)\epsilon{\hat\mu}\over2} +
{(a+1)^2\epsilon^2{\hat\mu}^2\over4}} 
\left [ d{\hat t}^2 - d{\hat r}^2 - d{\hat z}^2
- (1 - \epsilon{\hat\mu})^2 {\hat r}^{2 - 2(a+1)^2\epsilon^2{\hat\mu}^2/4} 
d\theta^2 \right
]
\eqno (3.17)
$$
which is almost, but not quite, a conformally rescaled cone. Note [14] that the
radius at which non-conical effects become important is when $R\simeq
e^{4\over(a+1)^2\epsilon^2{\hat\mu}^2}$ or $r\simeq \sqrt{\lambda}\eta
e^{4\over(a+1)^2\epsilon^2{\hat\mu}^2}$, therefore, 
for a typical GUT string, $r =
O(10^{100{\rm billion}})$! i.e.\ well beyond any reasonable cosmological scale.

This is reminiscent of metric of the global string[18], a system which has very
strong asymptotic effects and was for some time thought to be singular [19]. The
effect of the global string also becomes evident at very large radii
($e^{1/\epsilon}$), however, unlike the metric (3.16) the global string metric
is actually non-static and has an event horizon at finite distance from the
core [20].

Note that the back-reaction of the linearized solution (3.16) on the vortex
fields is to alter the long-range fall-off of the $X$ and $P$ fields:
$$
\eqalign{
1 - X &\simeq \exp \{ -R^{1+(a+1)\epsilon{\hat\mu}/4} \} \cr
P &\simeq \exp\{-R^{1+(a+1)\epsilon{\hat\mu}/4} /\sqrt{\beta} \} \cr
}
\eqno(3.18)
$$
which could be interpreted as a thickening of the core by a factor
$(1+(a+1)\epsilon{\hat\mu}/4)$.

Now let us consider the special case $a = -1$. In this case we see that
(setting $\gamma=\phi=0$ at the core) $\gamma = -2\phi$,
and (3.10c) implies that $\gamma'\to0$ rapidly outside the core.
In this case we see that to leading order, $\gamma$ takes its
Einstein form, and the back-reaction of the dilaton on the vortex fields 
serves only to perturb slightly the solution for the Einstein 
self-gravitating vortex. Thus for $a = -1$, the cosmic string is essentially 
the same as its Einstein gravity cousin. It therefore has the metric 
(2.17) and  $e^{2\phi} = e ^{-\gamma} \to e^{-C}$ which gives in the
string frame
$$
\eqalign{
d{\hat s}^2 &= dt^2 - dr^2 - dz^2 - 
{\tilde \alpha}^2_E e^{-2\gamma_E} d\theta^2 \cr
&\sim_{r\to\infty} dt^2 - dr^2 - dz^2 - (1 - \epsilon{\hat\mu})^2 
r^2 d\theta^2 \cr
}
\eqno (3.19)
$$
i.e.\ there is no red/blue-shift of time between the core and infinity in the
string frame, no matter what the value of $\beta$.

Finally, let us consider $\beta=1$. In this case, to linear order ${\cal P}_R =
{\cal P}_\theta=0$, and $\gamma=\phi=0$, which we suspect to be the case to all
orders, and indeed, the Bogomolnyi system (2.15) can be shown to provide the
solution to the fully self-gravitating string in this case.

Before moving on to the massive dilaton, it is worth emphasising that the 
$a = -1$ massless dilatonic cosmic string has no long range effects
(other than the deficit angle), and merely shifts the value of the dilaton
between the core and infinity by a constant of order $\epsilon$. For the
special case $\beta=1$, there is no effect at all on the dilaton field.

\vskip 3mm

\noindent{\it 3.2 Massive dilatonic gravity}

\vskip 2mm

In the absence of a preferred potential to take for the dilaton, we will use
${\tilde V}(\phi) = 2M^2\phi^2$, where $M = m/\sqrt{\lambda}\eta$ is the ratio
of the dilaton mass  to Higgs mass. Of course, we do not expect that this
will be the exact form of the dilaton potential, however, a quadratic
approximation will be valid provided $\phi$ remains close to the minimum of
the potential. For a GUT string we expect $10^{-11}\leq M\leq 1$, representing a
range for the unknown dilaton mass of 1TeV - $10^{15}$GeV. The dilaton equation
(3.10d), then becomes 
$$ 
(\alpha \phi')' = \alpha e^\gamma M^2 \phi + \epsilon
{(a+1)\over2} \alpha e^\gamma{\cal E} - {\half}\epsilon \alpha e^\gamma ({\cal
P}_R + {\cal P}_\theta)  
\eqno (3.20)
$$

Once again, we begin by considering the wire model for the string which again
gives $\alpha(R)$ and $\gamma(R)$ as in (3.13a,b). However, the presence of the
mass term in (3.20) now alters the form of the dilaton; integrating (3.20) for
the wire model gives 
$$
\phi_w =-{\half}(a+1)\epsilon{\hat\mu} K_0(MR)
\eqno(3.21)
$$
where $K_0$ is the modified Bessel
function. In this case, the constraint equation (3.10c) is satisfied for $R>M$,
but for $R<M$ we once again require $O(\epsilon^2)$ corrections, this is not
really surprising since this is within the Compton radius of the dilaton
and we might expect a behaviour analogous to that of the massless
dilaton. However, since these corrections are only significant for $R\simeq
e^{1/\epsilon^2}$, we can in this case safely ignore them. At the string
boundary, we have that $\phi \sim {\half}(a+1)\epsilon{\hat\mu}\ln M 
=$O($\epsilon$),
hence the quadratic approximation for the potential appears to be justified.

For an extended source, we may solve (3.20) implicitly using its Green's
function: 
$$
\eqalign{
\phi &= - {\half} \epsilon K_0(MR) \int_0^R I_0(MR') R' 
\left [ (a+1){\cal E}(R')
- ({\cal P}_R(R') + {\cal P}_\theta(R')) \right ] dR' \cr
& \;\;\; - {\half} \epsilon I_0(MR) \int_R^\infty K_0(MR') R' \left [ (a+1){\cal
E}(R') - ({\cal P}_R(R') + {\cal P}_\theta(R')) \right ]  dR' \cr 
&\simeq -{\half} (a+1)\epsilon {\hat\mu} K_0(MR) \;\;{\rm for}\;\;R>1, 
\;\; M\ll1 \cr
}
\eqno (3.22)
$$
which, unlike the massless dilaton case, is now in agreement with the wire model
estimate. A plot of $\phi(R)$ is illustrated in figure 2 for the
$\beta=1$ vortex shown in figure 1
with various values of $M$.
\midinsert \vskip -4truecm \hskip 2 truecm \epsfysize=15truecm
\epsfbox{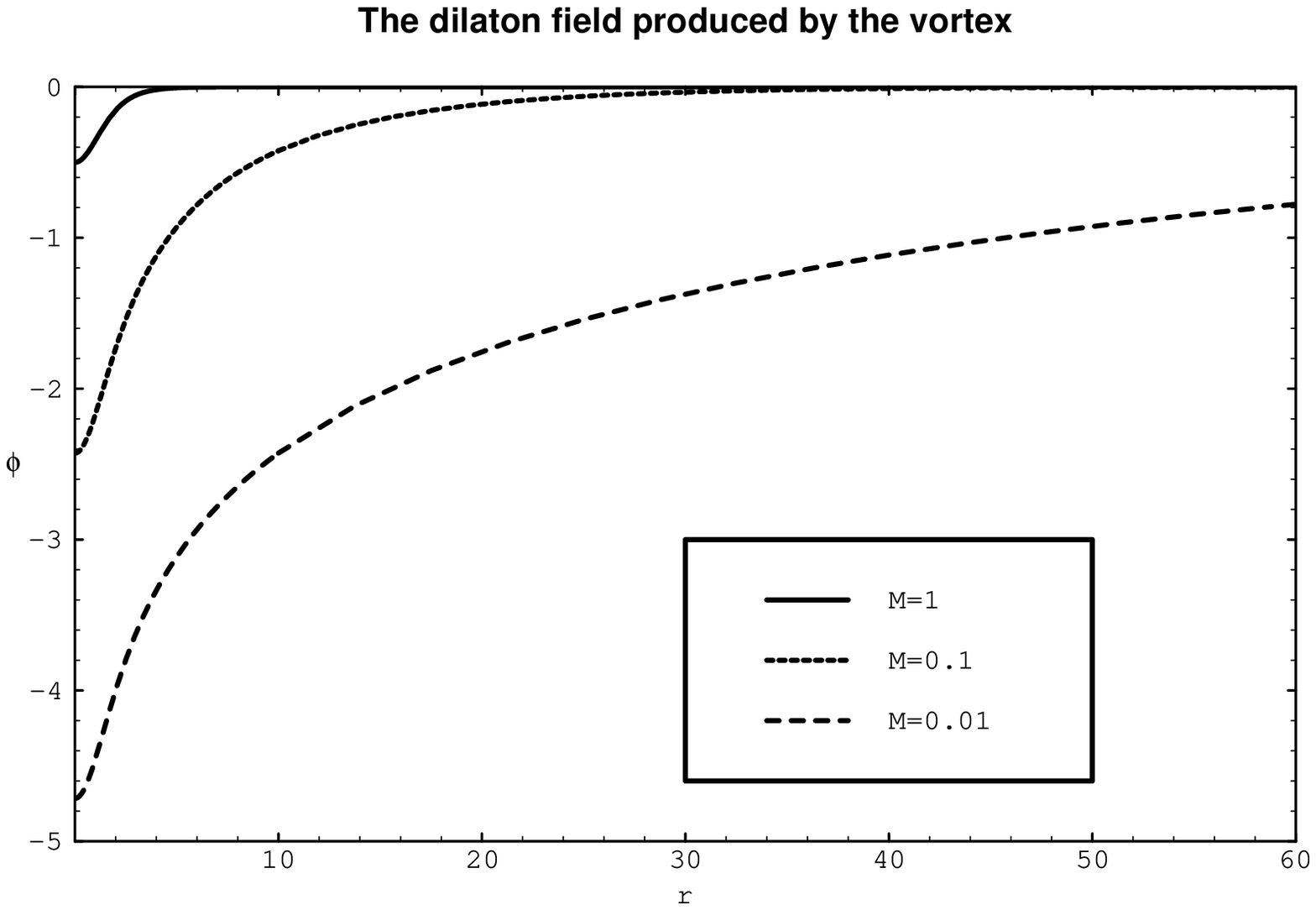} \vskip -3 truecm \hskip 2 truecm
\vbox{ \hsize=11.5 truecm
\noindent {\cat FIGURE (2): A plot of the dilaton field generated by
a $\beta=N=1$ vortex for various values of the dilaton mass. The factor
$(a+1)\epsilon{\hat \mu}$ has been scaled out of the dilaton. Note the
reciprocal dependence of the dilaton fall-off on the mass, compared
to the logarithmic dependence of the amplitude. }}
\endinsert

We may now write down the asymptotic solution for the cosmic string to order
$\epsilon$ as:
$$
\eqalign{
ds^2 &= e^{\gamma_E} \left [ dt^2 - dr^2 - dz^2 \right ] - \alpha^2_E
e^{-\gamma_E} d\theta^2 \cr
e^{2\phi} &= e^{-(a+1)\epsilon{\hat\mu} K_0(mr)} \cr
}
\eqno (3.23)
$$
Thus the spacetime is asymptotically conical in both string and Einstein frames.

Now consider $a = -1$. In this case the dilaton is very strongly damped to 
zero outside the core:
$$
\phi\simeq {\epsilon M\over 2} \left [ I_0(MR) \int_R^\infty K_0'(MR) R^2 {\cal
P}_R - K_0(MR) \int_0^R I_0'(MR)R^2{\cal P}_R \right ]
\eqno (3.24)
$$
therefore to a good approximation  $\phi=0$ outside the core, irrespective of
$M$, and therefore in both the Einstein and the string frames there is a red or
blue shift between the core and infinity.

Finally, if $\beta=1$, we once again have $\gamma=\phi=0$, and (2.15) gives the
first order equations of motion which this system satisfies.

\vskip 5mm

\noindent{\bf 4. Bogomolnyi bounds for dilatonic cosmic strings}

\vskip 2mm

The results of the previous section suggest that $a = -1$ is a rather
special point. Usually, for $\beta=1$, the Bogomolnyi limit, the equations
of motion for the cosmic string simplify -- they become first order -- and
the vortex saturates an energy bound determined by the winding number of the
vortex [21]. For the dilatonic vortex, this delicate balance appears
to be destroyed, except in the special case $ a = -1$. In this section
we would like to formalise this by presenting an energetic argument that
a topological bound can be saturated if and only if $ \beta = -a = 1$.

Since the cosmic string is cylindrically symmetric, and we do not 
{\it a priori} wish to make any assumptions about the global behaviour
of the spacetime, we use an energy tailored to the system at hand --
the C-energy introduced by Thorne[16]:
$$
E_c = 4\pi \left [ \gamma - \ln {\partial{\tilde\alpha}\over
\partial r} \right ] = 4\pi \left [ \gamma - \ln\alpha' \right ]
\eqno (4.1)
$$
modified slightly to allow for the absence of the Newton constant, $G$.
This energy can in turn be represented as the integral of the zeroth
component of a covariantly conserved C-momentum vector:
$$
E_c = \int {\tilde\alpha} e^\gamma P^0 dr
=\int \alpha e^\gamma {\hat P}^0 dR
\eqno (4.2)
$$
where
$$
\eqalign{
{\hat P}^0 &= {1\over 2\pi\alpha e^\gamma}
{\partial E_c \over \partial R} = {2 \over \alpha e^\gamma}
\left [ \gamma' - {\alpha''\over \alpha'} \right ] \cr
&= {2\over \alpha'} \left [ \epsilon {\cal E}
+ {\quarter} \gamma^{\prime 2} e^{-\gamma} + e^{-\gamma} \phi^{\prime2}
+
{\half} {\tilde V}(\phi) \right ] \cr
}
\eqno (4.3)
$$
Clearly every term in ${\hat P}^0$ is positive semi-definite, 
and all vanish only 
in flat space, the latter three vanishing if $\phi=\gamma=0$. Now consider
$\cal E$, we may rewrite this as
$$
\eqalign{
{\cal E} = e^{2(a+1)\phi} \Biggl \{ e^{-\gamma}
&\left [X'-e^\gamma{XP\over\alpha} \right ]^2
+\left [{P'\over\alpha}e^{-\phi} - {\half} e^\phi (X^2-1)\right ]^2\cr
&+ (\beta-1) {P^{\prime2}\over\alpha^2}e^{-2\phi} 
+ {1\over\alpha} \left [ (X^2-1)P\right ]' \Biggr \}\cr
}
\eqno (4.4)
$$
In order to get a `topological' value for the C-energy, we need
$(\gamma - \ln\alpha')$ to be expressed in terms of $X$ and $P$;
alternatively, we require ${\hat P}^0$ to be a total derivative.
For $a = -1$, $\beta = 1$, $\phi=\gamma=0$ and (2.15) implies that
all terms in ${\hat P}^0$ vanish except for the last expression in
equation (4.4) for ${\cal E}$. We thus obtain
$$
\eqalign{
E_c &= \int {2\epsilon\over \alpha'} 
\left [ (X^2-1)P\right ]' \cr
&= -2 \int \left ( \ln \left [ 1 - \epsilon[(X^2-1)P+1] \right ] \right )' \cr
&= -2\ln(1-\epsilon) = \epsilon + .....
= \eta^2 + O(\eta^4) \cr
}
\eqno (4.5)
$$
--the topological bound. For a string with winding number other than one, 
we replace $P$ by $NP$ and hence $E_c$ becomes $-2\ln(1-N\epsilon)=
N\eta^2 + O(\eta^4)$.

For $\beta\neq1$ it is immediately clear that this topological bound
cannot be saturated due to the presence of the $(\beta-1)P^{\prime2}$
term in the integral. Similarly, if $a\neq-1$, the equation of motion
for $\phi$ shows that $\phi'$ must be nonzero due to the presence of
the $(a+1){\cal E}$ term on the right hand side of (3.10d), hence 
${\hat P}^0$ is strictly greater than $2\epsilon{\cal E}/\alpha'$,
and once again, the topological bound cannot be saturated.

Therefore, by considering a fully covariant relativistic definition
of energy for cylindrically symmetric systems, we have shown that there
exists a topological `bound' for the energy of the vortices, in a rather
analogous fashion to the topological quantity originally derived by 
Bogomolnyi[15] for flat space vortices, and this bound is saturated
only for $\beta=-a=1$.

\vskip 5mm
\noindent{\bf 5. Geodesics}

In this section we discuss the motion of test particles
following geodesics in the spacetimes presented in section three.
According to experimental tests [22] any theory describing gravity has to 
verify the Weak Equivalence Principle (WEP).
This principle states that any path through spacetime
of a freely falling neutral test body is independent
of its structure and composition.
Therefore gravity has to couple in the same way
to massive test particles and to photons.  The obvious way is coupling 
directly to the metric in the string frame,  which is what 
one usually does in scalar-tensor theories.
Clearly, since the string and Einstein frames are related by
a conformal transformation, null geodesics will be the same in either frame,
but the geodesics of massive particles will be different. 

We begin by commenting on the massive dilaton.
Here the metric is given by (2.17) outside the Compton radius of the
dilaton, and is therefore conical. Geodesics are therefore the same as
for the Einstein cosmic string, and indeed, since the corrections within 
the Compton radius of the dilaton are extremely small (O($\epsilon^2$)),
the geodesics throughout the whole spacetime in the Einstein frame are
essentially the same as for the Einstein self-gravitating string.

Now consider the massless dilaton.
In the string frame the metric is given by eq.(3.17)
and the radial motion of a test particle in a plane transverse to the string,
$d{\hat z}=0$, is given by:
$$
{\dot {\hat r}}^2 + {h^2 \over {\hat r}^{2(1+\nu)}} +
{k \over {{\hat r}^{\nu(1+\nu)}}}
= {E^2 \over {{\hat r}^{2\nu(\nu+1)}}}
\eqno (5.1)
$$ 
where $\nu = (a+1)\epsilon{\hat\mu}/2$, and
the dot denotes a derivative with respect to the proper time
along a timelike geodesic, or an affine parameter for photons.
The parameter $k$ is either one or zero, representing either a
massive particle or photon respectively. $E$ and $h$ are constants 
of the motion  representing energy and angular momentum respectively, 
and are given by:
$$
\eqalign{
E &= {\hat g}_{tt} {{\dot {\hat t}}} = {\hat r}^{\nu(\nu+1)} {\dot{\hat t}}\cr
h &= (1-\epsilon{\hat\mu}){\hat g}_{\theta\theta} {{\dot\theta}}
= (1 - \epsilon {\hat \mu})^3 {\hat r}^{2-\nu^2+\nu} {\dot\theta} \cr
}
\eqno (5.2)
$$

For  $a=-1$, $\nu = 0$, and
irrespective of whether the dilaton is massive or massless, the geodesics
are qualitatively the same as for the Einstein cosmic string. Indeed, from
(5.1) one sees that the radial motion of a geodesic is the same as the
classical trajectory of a unit mass particle of energy ${E^2 \over 2}$,
with an effective potential given by:
$$
V_{_{\rm eff}} = {h^2 \over 2{\hat r}^2} + {k \over 2}
\eqno (5.3)
$$
which is an identical effective potential to that of a particle moving in
flat space. (The presence of the $\epsilon{\hat\mu}$ terms in the definition
of $h$ shows that the spacetime is not globally flat, but conical.)
All non-static trajectories therefore escape to infinity, and satisfy
$$ 
{\hat r} \geq {h \over {\sqrt{E^2-k}}}
\eqno (5.4)
$$
In addition, there exist static trajectories for massive particles:
${\hat r} = {\hat r}_0$, $E=1$.

Now consider $a \not= -1$.
For comparison with the effective potential (5.3), it is useful to 
redefine the radial coordinate ${\hat r}$ via
$$
\rho = {{{\hat r}^{\nu^2+\nu+1}} \over {\nu^2+\nu+1}}
\eqno (5.5)
$$
which gives the $\rho$-radial motion as that of a 
unit mass particle of energy ${{E^2} \over 2}$, 
with an effective potential given by:
$$
U_{_{\rm eff}} = {h^2\over 2[(\nu^2+\nu+1)\rho]^{2(1-\nu^2)\over \nu^2+\nu+1}}
+ {k\over2}[(\nu^2+\nu+1)\rho]^{\nu(\nu+1)\over  \nu^2+\nu+1}
\eqno (5.6)
$$
Since $\nu = O(\epsilon)$, to leading order this is
$$
U_{_{\rm eff}} \simeq {h^2\over 2(1+2\nu) \rho^{2(1-\nu)}} +
{k\over2} \rho^\nu.
\eqno(5.7)
$$

First consider $\nu>0$, i.e.\ $a>-1$.
For massless particles, $U_{_{\rm eff}}\simeq V_{_{\rm eff}}$, and thus
photons escape to infinity, however, note that as ${\hat r}\to\infty$,
$g_{tt} = {\hat r}^{\nu(\nu+1)}\to\infty$, hence these photons will be
infinitely redshifted. (Note that this will happen in either frame, although
the red-shifting in the Einstein frame occurs at a rate proportional to
$\epsilon^2$ rather than $\epsilon$.)
\midinsert \vskip -5truecm \hskip 1truecm \epsfysize=18truecm
\epsfbox{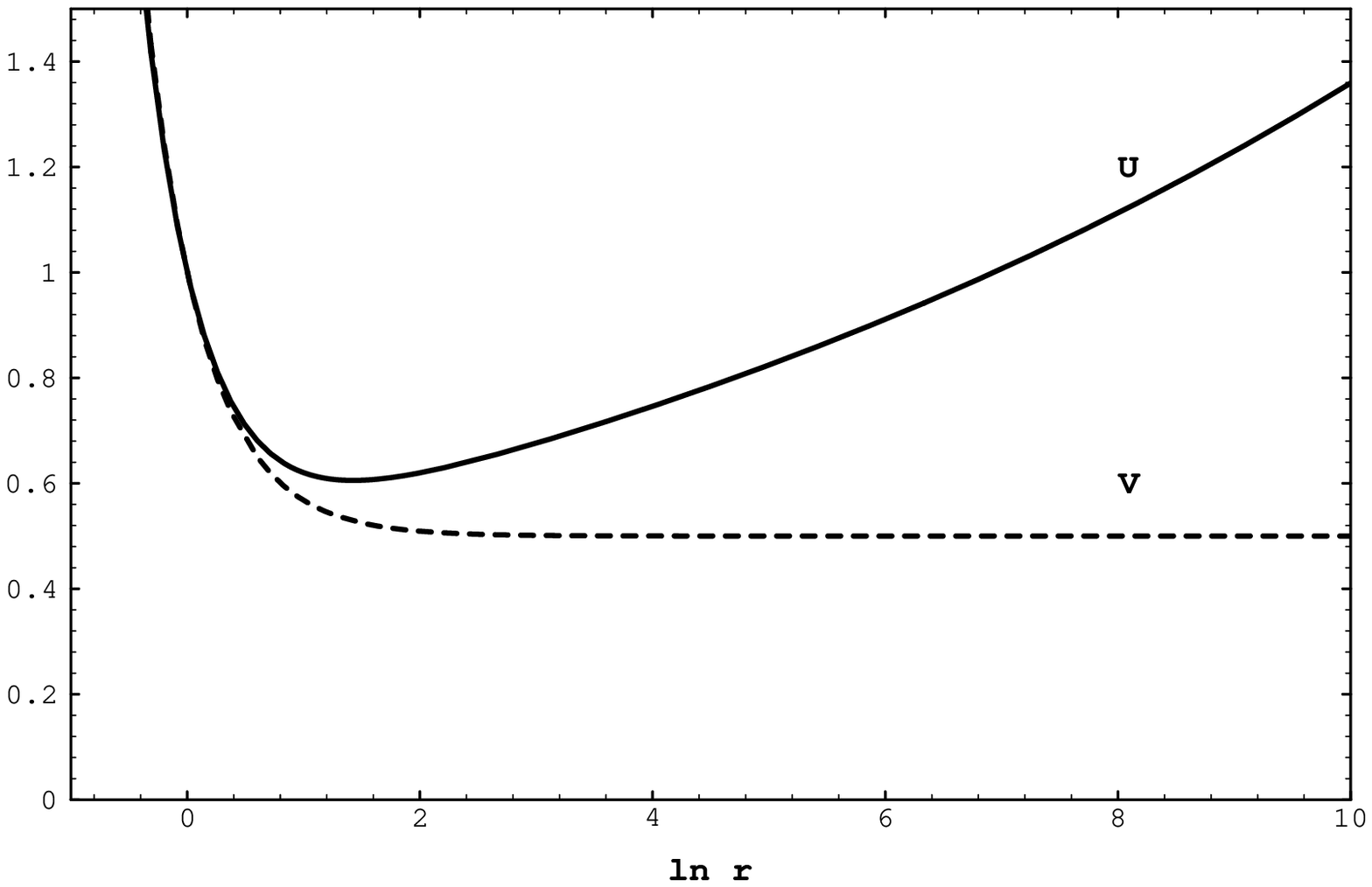} \vskip -5 truecm \hskip 2 truecm
\vbox{ \hsize=11.5 truecm
\noindent {{\cat FIGURE (3): A comparison of the effective potential
for a massive particle in a conical (dashed) and the massless
dilaton (solid) background for $\nu =0.1, h =1$. } }}
\endinsert

For massive particles, $U_{_{\rm eff}}$ is now a potential well, 
(see figure 3) hence
{\it all} trajectories of massive particles are bounded, however,
for $\rho \ll e^{1/\nu}$, $U_{_{\rm eff}}\simeq V_{_{\rm eff}}$ hence
trajectories approaching `close' to the cosmic string (i.e.\ on
all scales of cosmological interest) behave as if in a conical
spacetime. Such orbits will be highly eccentric, and have an outer 
bound of ${\hat r} = O( E^{1/\nu})$.
Note that there are no static geodesics in this case,
all particles initially at rest will be attracted to the string by an
acceleration of order $\epsilon/r_0$.

If $\nu<0$, i.e.\ $a <-1$, then $U_{_{\rm eff}}$ is once more a
scattering potential  and all particles escape to infinity. Since
$g_{tt} \to 0$ in this case, photons will now be infinitely blue-shifted.
Once again there are no static solutions to the massive particle
geodesic equation, this time the particles are repelled from the string.

In the Einstein frame, the photon trajectories are identical to
that of the string frame, but now all the massive particle trajectories
are bound, as can be seen by removing all the terms involving $\nu$
from (5.6).

\vskip 5mm

\noindent{\bf 6. Discussion}

\vskip 2mm

In this paper, we have derived the metric for U(1) local cosmic strings
in dilaton gravity both with and without a potential for the dilaton.
The (unknown) coupling of the abelian-Higgs model to the dilaton is
accounted for by coupling the Lagrangian to the gravitational sector by an
arbitrary $e^{2a\phi}$ factor. 

For a massless dilaton, the results are qualitatively the same as those of 
Gundlach and Ortiz [14], who considered cosmic strings in JBD theory. 
Essentially, the metric is the same as the usual cosmic string, i.e.\ conical,
in the Einstein frame, and conformally conical in the string frame on scales of
cosmological interest. However, on the very large scale, ($r\sim
\sqrt{\lambda}\eta e^{4\over (a+1)^2 \epsilon^2{\hat\mu}^2}$), there
is additional curvature, and the spacetime is not asymptotically locally
flat in either frame. The exception is the special case $a = -1$,
in which the metric is conical in either frame, and the dilaton is shifted
in the core relative to infinity, the direction of the shift depending on
whether the cosmic string is type I or II, no alteration in the dilaton
occurring for the boundary between types I and II: $\beta=1$.

For a massive dilaton, as expected, the metric asymptotes a conical metric,
in both string and Einstein frames, however, the string does generate a
dilaton `cloud', approximately of width $m_H/m_\phi$, which is 
schematically depicted in figure 3, for $a \neq -1$. For $a = -1$ the
dilaton is only perturbed away from its vacuum value in the core
of the string, and for $\beta=1$, it is not affected at all.
\midinsert \hskip 4 truecm \epsfysize=10truecm
\epsfbox{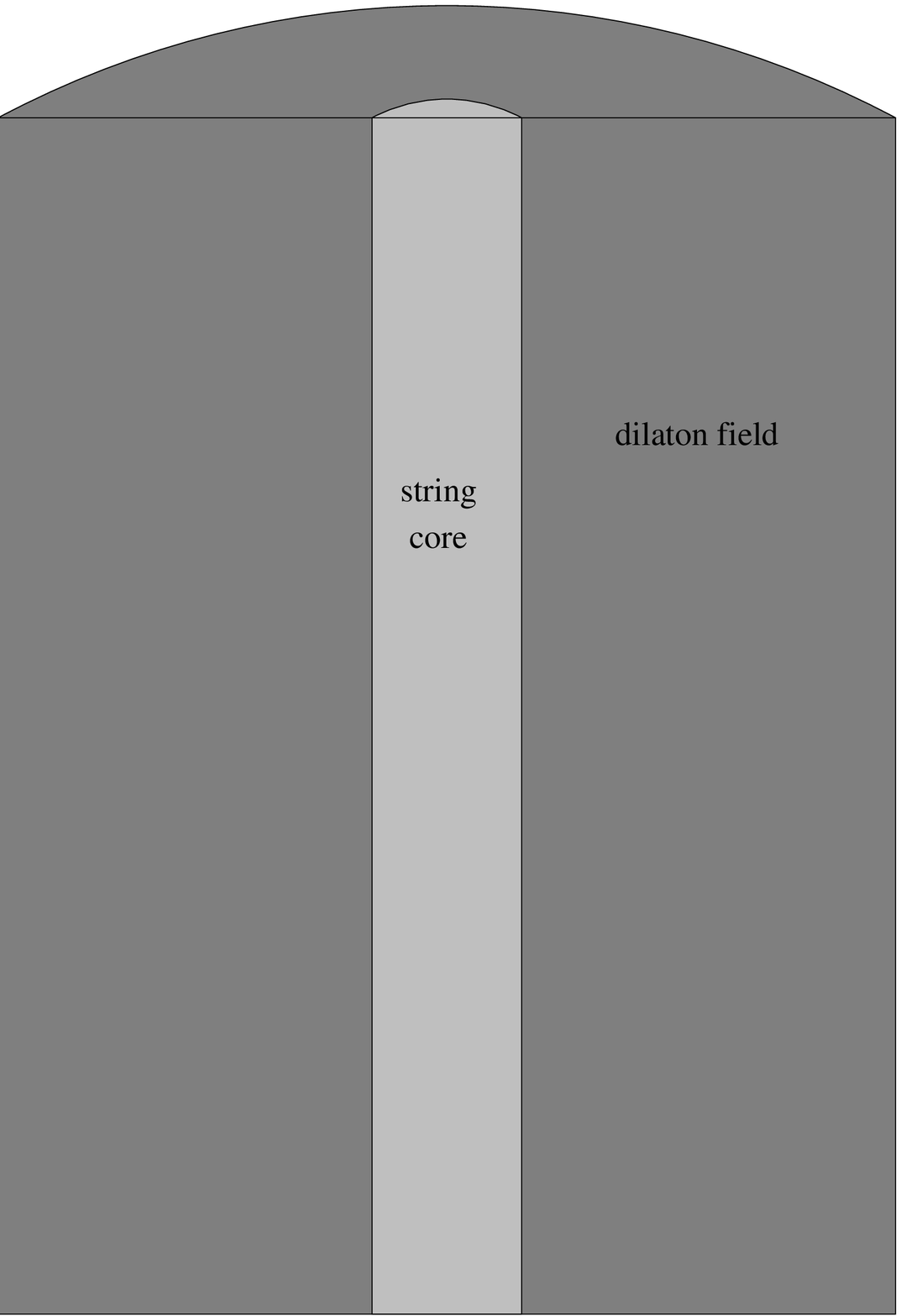} \medskip \hskip 2 truecm
\vbox{ \hsize=11.5 truecm
\noindent {\cat FIGURE (4): A representation of the dilaton field
surrounding the cosmic string for $ a \neq -1$. }}
\endinsert

Although it is beyond the scope of this paper to derive the effective
action of the cosmic string, the results do support a Nambu approximation
for the string, since they show that the metric is little affected on
cosmological length scales, and remains approximately flat locally
(unlike the global string [18]). Damour and Vilenkin [23] have recently
explored the impact of a massive dilaton on string networks
using a model for the interactions which modifies the Nambu approximation
by making the mass per unit length interact with the (massive) dilaton.
In other words, the worldsheets act as sources for the dilaton which has
a mass $m_\phi$. They concluded that a TeV mass dilaton was incompatible
with a GUT string network. Our results largely back up this calculation,
but with one important caveat: The model used by Damour and Vilenkin
makes no reference to the details of the dilaton coupling to the
particle physics model producing the strings, the abelian-Higgs
lagrangian, {\it i.e.\ their coupling is independent of our variable $a$}.
Therefore, one should renormalize their calculations by factors of 
$(a+1)$. This means that the conclusion that a TeV mass dilaton is
incompatible with string theories of structure formation is only
valid if $a$ is not close to $-1$. For $a = -1$, such as might be the
case if the fields composing the string are derived from
heterotic string theory or the NS-NS sector of type II string 
theory for example, there will be little dilatonic radiation from the 
cosmic string network, and hence a much weaker constraint.

To sum up: the gravitational field of a cosmic string in dilaton
gravity is surprisingly close to that of an Einstein cosmic string on
cosmological distance scales. However, it is the microwave background
rather than cosmological observations, that provides the tightest
constraint on the cosmic string theory of structure formation. If the
strings couple to the dilaton directly ($a = -1$), then such constraints
are identical to those derived in Einstein gravity. However, if the
string couples with $a$ different from $-1$, then the constraints
of Damour and Vilenkin [23] apply, and a `low' (i.e.\ close to electroweak)
mass for the dilaton rules out the cosmic string scenario of galaxy
formation.

\vskip 5mm

\noindent{\bf Acknowledgements.}

\vskip 2mm

It is a pleasure to thank Filipe Bonjour for helpful discussions. 
This work was supported by a Royal Society University
Research Fellowship (R.G.), and a JNICT fellowship BD/5814/95 (C.S.).

\vskip 5mm

\noindent{ \bf References.}

\vskip 2mm

\nref
R.H.Brandenberger, {\it Modern Cosmology and Structure Formation}
astro-ph/9411049.

M.B.Hindmarsh and T.W.B.Kibble, \rprog 58 477 1995. [hep-ph/9411342]

A.Vilenkin and E.P.S.Shellard, {\it Cosmic strings and other 
Topological Defects} (Cambridge Univ. Press, Cambridge, 1994).
 
\nref
A.Vilenkin, \prd 23 852 1981.

J.R.Gott III, \apj 288 422 1985.

W.Hiscock, \prd 31 3288 1985.

B.Linet, \grg 17 1109 1985.
 
\nref
D.Garfinkle, \prd 32 1323 1985.

\nref
R.Gregory, \prl 59 740 1987.
 
\nref
E.Fradkin, \plb 158 316 1985.

C.Callan, D.Friedan, E.Martinec and M.Perry, \npb 262 593 1985.

C.Lovelace, \npb 273 413 1985.

\nref
P.Jordan, \zphys 157 112 1959.\hfill\break
C.Brans and R.H.Dicke, \pr 124 925 1961.

\nref
G.Veneziano, \plb 265 287 1991. \hfill\break
A.A.Tseytlin and C.Vafa, \npb 372 443 1992. [hep-th/9109048] \hfill\break
A.Tseytlin, \ijmpd 1 223 1992. [hep-th/9203033] \hfill\break
D.Goldwirth and M.Perry, \prd 49 5019 1994. [hep-th/9308023]\hfill\break
E.Copeland, A.Lahiri and D.Wands, \prd 50 4868 1994. [hep-th/9406216]

\nref
J.D.Barrow and K.Maeda, \npb 341 294 1990.\hfill\break
A.Burd and A.Coley, \plb 267 330 1991.\hfill\break
J.D.Barrow, \prd 47 5329 1993. \prd 48 3592 1993.\hfill\break
A.Serna and J.Alimi, \prd 53 3074 1996. [astro-ph/9510139]\hfill\break
S.Kolitch and D.Eardley, \annph 241 128 1995.\hfill\break
C.Santos and R.Gregory, {\it Cosmology in Brans-Dicke theory with a scalar
potential}, gr-qc/9611065.

\nref
N.Turok, \prl 63 2625 1989.

\nref
A.Albrecht, D.Coulson, P.Ferreira and J.Magueijo, \prl 76 1413 1996.
[astro-ph/9505030] \hfill\break
N.Crittenden and N.Turok, \prl 75 2642 1995. [astro-ph/9505120]\hfill\break
R.Durrer, A.Gangui and M.Sakellariadou, \prl 76 579 1996. [astro-ph/9507035]

\nref
H.B.Nielsen and P.Olesen, \npb 61 45 1973.

\nref
D.Forster, \npb 81 84 1974.\hfill\break
K.I.Maeda and N.Turok, \plb 202 376 1988.\hfill\break
R.Gregory, \plb 206 199 1988. \prd 43 520 1991.

\nref
R.Geroch and J.Traschen, \prd 36 1017 1987..

\nref
C.Gundlach and M.Ortiz, \prd 42 2521 1990. \hfill\break
L.O.Pimental and A.N.Morales, {\it Rev.\ Mex.\ Fis.\ }{\bf 36} S199 (1990).
\hfill\break
M.E.X.Guimaraes, gr-qc/9610007.

\nref
E.B.Bogomolnyi, {\it Yad.~Fiz.} {\bf24} 861 (1976)[{\it
Sov.~J.~Nucl.~Phys.~{\bf 24}} 449 (1976)]
 
\nref
K.S.Thorne, \pr 138 251 1965.
 
\nref
T.Levi-Civita, {\it Atti Acc.\ Lincei.\ Rend.\ }{\bf 28} 101 (1919).

\nref
A.G.Cohen and D.B.Kaplan, \plb 215 67 1988.

\nref
R.Gregory, \plb 215 663 1988.\hfill\break
G.Gibbons, M.Ortiz and F.Ruiz, \prd 39 1546 1989.
 
\nref
R.Gregory, \prd 54 4955 1996. [gr-qc/9606002]

\nref
B.Linet, \pla 124 240 1987. \hfill\break
A.Comtet and G.Gibbons, \npb 299 719 1988.

\nref
R.E\"otv\"os, V. Pek\'ar, E. Fekete, \anndph 68 11 1922. \hfill\break
R.H.Dicke, \amjph 28 344 1960. \hfill\break
V.B.Braginsky, V.I. Panov, \jtp 34 463 1972. 

\nref
T.Damour and A.Vilenkin, {\it Cosmic strings and the string dilaton},
gr-qc/9609067.

\bye